\documentclass[12pt,onecolumn]{emulateapj}

\usepackage{amsmath}

\newfont{\myfont}{cmmib10}

\newfont{\myfontsmall}{cmmib8}

\DeclareSymbolFont{cmmi}{OML}{cmm}{m}{it}
\DeclareMathSymbol{v}{\mathalpha}{cmmi}{"76}

\shorttitle{Temporal Broadening of Transient Radio Emission by the IGM}
\shortauthors{Macquart \& Koay}

\begin{document}

\title{Temporal Smearing of Transient Radio Sources by the Intergalactic Medium}


\author{Jean-Pierre Macquart\altaffilmark{1,2} 
and Jun Yi Koay\altaffilmark{1,3}
}

\altaffiltext{1}{ICRAR/Curtin University, Curtin Institute of Radio Astronomy, Perth WA 6845, Australia {\it J.Macquart@curtin.edu.au}}
\altaffiltext{2}{ARC Centre of Excellence for All-Sky Astrophysics (CAASTRO)}
\altaffiltext{3}{Department of Internet Engineering and Computer Science, Universiti Tunku Abdul Rahman, Setapak, KL 53300, Malaysia}

\begin{abstract}
The temporal smearing of impulsive radio events at cosmological redshifts probes the properties of the ionized Inter-Galactic Medium (IGM).  We relate the degree of temporal smearing and the profile of a scattered source to the evolution of turbulent structure in the IGM as a function of redshift.  We estimate the degree of scattering expected by analysing the contributions to the Scattering Measure (SM) of the various components of baryonic matter embedded in the IGM, including the diffuse IGM, intervening galaxies and intracluster gas.  These estimates predict that the amount of temporal smearing expected at 300\,MHz is typically as low as $\sim 1\,$ms and suggests that these bursts may be detectable with low frequency widefield arrays.  A generalization of the DM-SM relation observed for Galactic scattering to the densities and turbulent conditions relevant to the IGM suggests that scattering measures of order $10^{-6}$\,kpc\,m$^{-20/3}$ would be expected at $z \sim 1$.  This scattering is sufficiently low that its effects would, for most lines of sight, not be manifest in existing observations of the scattering broadening in images of extragalactic compact sources.  The redshift dependence of the temporal smearing discriminates between scattering that occurs in the host galaxy of the burst and the IGM, with $\tau_{\rm host} \propto (1+z)^{-3}$ if the scattering probes length scales below the inner scale of the turbulence or $\tau_{\rm host} \propto (1+z)^{-17/5}$ if the turbulence follows a Kolmogorov spectrum. This differs strongly from the expected IGM scaling $\tau_{\rm IGM} \sim z^2$ for $z \la 1$ and $(1+z)^{0.2-0.5}$ for $z \ga 1$. 
\end{abstract}


\section{Introduction}
There have been six recent detections of transient radio bursts whose dispersion measures (DMs) are so large that their emission has been suggested to originate at cosmological distances (Lorimer et al.\,2007; Keane et al.\,2010; Thornton et al.\,2013).  A substantial fraction of the DMs of these bursts are attributed to baryons in the diffuse Inter-Galactic Medium, heralding the possibility of probing the entire baryonic content of the Universe out to which these bursts can be detected.  Measurement of the dispersion delay as a function of frequency enables direct measurement of the total electron column along the line of sight, and for $z \ga 0.2$ the contribution of the IGM is expected to dominate the total DM of objects viewed through lines of sight off the Galactic plane (Ioka 2003; Inoue 2004).  

The existence of such bright, short-duration transients makes it possible to probe the ionized IGM at cosmological distances in exquisite detail.  Impulsive extragalactic radio bursts offer the prospect of measuring the turbulent properties of the Inter-Galactic Medium (IGM).  Turbulent plasma along the line of sight can scatter the radiation, which alters the temporal profile of the burst as it propagates through intergalactic plasma.  The effect of temporal smearing caused by scattering potentially reveals information on the IGM on exquisitely fine scales, down to lengths as small as $\sim 10^{10}\,$m.   
Temporal smearing arises due to multipath propagation as the radiation travels through  inhomogeneities in the turbulent plasma.  Radiation scattered back into the line of sight is delayed relative to radiation that arrives along the direct line of sight to a source, with the amount of radiation scattered depending on the typical angle through which it is scattered.  This angle, and hence the amount of delayed radiation, increases with the strength of the scattering.  Thus both the apparent angular diameter and the temporal profile of a compact transient source depend on the amount and distribution of turbulent intergalactic plasma along the line of sight.  

There is strong evidence that the radiation from these extragalactic bursts is subject to scattering at intergalactic distances.  In two out of the six reported cases the burst duration increases sharply with wavelength across the observing band, scaling as $\lambda^{4.0 \pm 0.4}$ (Lorimer et al.\,2007; Thornton et al.\,2013).  This scaling is characteristic of radiation subject to temporal broadening caused by turbulent plasma.  The reported smearing times are of order milliseconds, much larger than the microsecond timescales expected of temporal smearing expected due to the turbulence in our own Galaxy along the lines of sight on which these bursts are observed.  The expected smearing time of the Lorimer burst, if due to Galactic scattering alone, is $\sim 0.5\,\mu$s, far smaller than the observed width of $\sim 5\,$ms.  This raises the prospect that radiation from these bursts contains information on turbulence in the ionized IGM and even in the host galaxy.

There are two main reasons to compute the effect of temporal smearing due to the IGM.  Firstly, it affects searches for short-duration transients at sufficiently low frequency.  The steep scale of the temporal smearing timescale with frequencies raises the possibility that searches for transients at low frequencies are particularly susceptible to the effects of temporal smearing.  Unlike dispersion smearing, temporal smearing due to multipath propagation results in an irretrievable loss of sensitivity to transients when the smearing time exceeds the intrinsic burst duration, with the observed S/N decreasing by the square-root of the factor over which the smearing time exceeds the intrinsic pulse duration.

Secondly, at higher frequencies, where temporal smearing is a smaller hindrance to the detection of transient sources but is still measurable, the effect affords a means of probing the evolution of the IGM itself.  This is attractive because it furnishes a means of probing the history of energy deposition in the IGM.  The sources invoked to explain reionization of the Universe and feedback associated with galaxy formation inject energy into the IGM on large scales, driving a turbulent cascade that should drive the evolution of inhomogeneity in the IGM.  Sources of energy input include the radiative and mechanical energy of AGN and their jets, the UV and X-ray emission and stellar winds from young stars and the flows driven by supernovae (Cen \& Ostriker 2006 and references therein).  



The subject of intergalactic scattering is coming to the fore with the inevitability of detecting many more extragalactic transients in the near future.  Several radio arrays with wide fields of view ($\ga 10$\,sq.\,deg.) ideally suited to the detection of large numbers of short-duration transient sources are currently entering operation or being planned.  These telescopes span the frequency range 30\,MHz to 3\,GHz and include LOFAR, the Murchison Widefield Array (MWA; Tingay et al. 2013), ASKAP (Johnston et al. 2007), MeerKAT and, eventually, the SKA.  Specific surveys for fast transients are planned on many of these, including the LOFAR Transients Key Project (Stappers et al.\, 2011), the Commensal Real-time ASKAP Fast Transients survey (CRAFT; Macquart et al.\,2010), and the Transients and Pulsars with MeerKAT survey (TRAPUM; PIs Stappers and Kramer).

The Parkes High Time Resolution Universe Legacy survey (HTRU; Keith et al.\,2010), even with a comparatively small field of view, has shown that the rate of impulsive radio extragalactic transient events is large.  The detection of the four high-DM pulses so far implies an event rate $\sim 10000_{-5000}^{6000}\,$sky$^{-1}$day$^{-1}$ at 1.4\,GHz above 1\,Jy (Thornton et al. 2013.).  The high luminosity of the events detected so far from inferred redshifts of $0.2 \la z \la 0.6$ suggests that it is relatively easy to detect bursts out to redshifts with lookback times to a significant fraction of the age of the Universe.  This rate may be augmented from other types of yet-undetected events, such as those postulated to be generated by coherent radio emission associated with the initial explosions of GRBs (Usov \& Katz 2000, Sagiv \& Waxman 2002, Moortgat \& Kuijpers 2004).  

Should they be detected in great enough numbers, the combination of DM and scattering information gleaned from transients may enable tomographic reconstruction of the structure and turbulent properties of the baryonic component of the Universe.  This is akin to the manner in which pulsar measurements have been used to map out the structure of our Galaxy's own interstellar medium (e.g. Armstrong, Rickett \& Spangler 1995; Taylor \& Cordes 1993; Cordes \& Lazio NE2001).

In this paper we relate the temporal smearing of radio transients to the underlying properties of the IGM.  In \S2 we outline the relationship between angular broadening and temporal smearing due to scattering to the turbulent properties of a plasma located at cosmological distances.  The aim of this section is to furnish the means by which future transients detections may be used to reverse engineer the structure of the IGM.  In \S3 we explore simple models for turbulence in the IGM and show how the magnitude of scattering effects would be expected to scale with redshift under several simple scenarios.  In \S4 we discuss existing limits and properties of IGM turbulence based on measurements to date.  We conclude in \S5 with a summary of the means by which these effects may be used as a cosmological tool to probe the history of energy deposition into the IGM.

\section{Scattering in curved spacetime}

We consider the effects of scattering under the simplifying assumption that the inhomogeneities associated with the plasma are confined to a single plane.  The thin-screen approximation provides an accurate description of the scattering properties if the line of sight is dominated by a single turbulent region.  However the range of validity of this commonly-used approximation extends further because it is often possible to model the effects of an extended medium in terms of an equivalent thin-screen after appropriate adjustment of simple parameters, such as the screen distance and scattering strength (see Tatarski \& Zavorotnyi (1980) and Lee \& Jokipii (1975) in the context of temporal smearing).  The thin-screen formalism elucidates the essential physics of the scattering without the hindrance of the extra mathematical formalism required to exhaustively treat a scattering in an extended medium.  


The observed wavefield of a point-like source of unit amplitude emanating at cosmological distances with angular diameter distance $D_S$ and subject to phase fluctuations on a plane located an angular diameter distance $D_L$ from an observer is (Schneider et al. 1992; see also Macquart 2004),
\begin{eqnarray}
u({\bf X}) &=&  \frac{e^{-i \pi/2} }{2 \pi r_{\rm F}^2} \int d^2 {\bf x} \exp \left[ \frac{i}{2 r_{\rm F}^2} \left( {\bf x}- \frac{D_{LS}}{D_{S}}{\bf X} \right)^2 +   i \phi({\bf x})  \right],  \label{uGrav} \\
r_{\rm F}^2 &=& \frac{D_L D_{LS} \lambda_0 }{2 \pi D_S (1 + z_L)}, \label{rF}
\end{eqnarray}
where $r_{\rm F}$ is the Fresnel scale, ${\bf X}$ is a co-ordinate in the plane of the observer, $z_L$ is the redshift of the scattering material, $\lambda_0$ is the wavelength in the observer's frame, $D_{LS} \neq D_S - D_L$ is the angular diameter distance from the source to the phase plane and $\phi$ represents the phase delays imparted to the radiation on this plane by the IGM.  Eq.\,(\ref{uGrav}) is the generalization of the Fresnel-Kirchoff integral to curved spacetime geometries; it retains the same form as in Euclidean space, but with the details that embody the curved geometry of the Universe contained within the calculation of the angular diameter distance\footnote{The angular diameter distance at a redshift $z$ is given by the integral $D(z) =c H_0^{-1} (1+z)^{-1} \int_0^z [\Omega_\Lambda + (1-\Omega) (1+z')^2 + \Omega_m (1+z')^3 + \Omega_r (1+z')^4 ]^{-1/2} dz' $, where $H_0$ is the Hubble constant, and $\Omega = \Omega_\Lambda + \Omega_m + \Omega_r$ and $\Omega_\Lambda$, $\Omega_m$ and $\Omega_r$ are, respectively, the ratios of the dark energy density, matter density and radiation density to the critical density of the Universe.  Throughout this paper we take $\Omega =1$ and $\Omega_r=0$.}.  In the formalism used in the treatment of gravitational lensing which, like the current situtation, involves optics at cosmological distances, the Fresnel-Kirchoff integral is often written in the alternate form which makes explicit the time delay of the radiation, $t_d({\bf x},{\bf X})$:
\begin{eqnarray}
u({\bf X}) =  \frac{e^{-i \pi/2} }{2 \pi r_{\rm F}^2} \int d^2 {\bf x} \exp \left[ 2 \pi i \nu t_d({\bf x},{\bf X})  \right],  \quad  t_{\rm d}({\bf x},{\bf X}) = \left[ \frac{D_S (1+z_L)}{2c D_L D_{LS}} \left( {\bf x} - \frac{D_{LS}}{D_S} {\bf X} \right)^2 + \frac{\phi({\bf x})}{2 \pi \nu} \right].
\end{eqnarray}

  
In order to compute quantities involving phase fluctuations, a model is needed to specify the electron density fluctuations.  In a wide variety of turbulent astrophysical plasmas the electron density power spectrum, $\Phi_{N_e}$, is taken to follow a power law between some inner and outers scales $l_0$ and $L_0$ respectively.  The amplitude of the turbulence per unit length, $C_N^2$, is parameterized in terms of distance along the ray path, $l$, so that the power spectrum takes the form,
\begin{eqnarray}
\Phi_{N_e} ({\bf q};l) = C_N^2(l) \, q^{-\beta} e^{-(q l_0)^2}, \qquad q > \frac{2 \pi}{L_0}. \label{PhiDefn}
\end{eqnarray}
The specific choice of index $\beta=11/3$ corresponds to the value associated with Kolmogorov turbulence; this index is approximately consistent with interplanetary and interstellar plasma measurements (Armstrong, Rickett \& Spangler 1995).  We confine our results to the regime $\beta < 4$, which pertains to most astrophysical plasmas.  

A useful related quantity is the phase structure function, which measures the square phase difference between two points separated by a displacement ${\bf r}$ on the phase screen,  
\begin{eqnarray}
D_\phi({\bf r}) &=& \langle [ \phi({\bf r}+{\bf r}') -\phi({\bf r}')]^2 \rangle \\ 
&=& \frac{2\,r_e^2 \lambda_0^2}{(1+z_L)^{2}} \int_l^{l+\Delta L} dl \int d^2{\bf q}\, (1-e^{i{\bf q} \cdot {\bf r}}) \Phi_{N_e}({\bf q},q_l=0;l), \label{DphiDefn}
\end{eqnarray}
where $\Delta L$ is the (small) thickness of the phase screen.  For a power law spectrum of density inhomogeneities it is convenient to write the phase structure function in the form $D_\phi (r) = (r/r_{\rm diff})^{\beta - 2}$ with 
\begin{subequations}
\begin{eqnarray}
r_{\rm diff} &=& \begin{cases} 
\left[ \frac{\pi r_e^2 \lambda_0^2}{(1+z_L)^2} \,  {\rm SM} \, l_0^{\beta-4} \, \frac{\beta}{4} \, \Gamma \left(-\frac{\beta}{2}\right)  \right]^{-1/2}, &  r_{\rm diff} < l_0, \\
\left[  2^{2-\beta} \frac{\pi r_e^2 \lambda_0^2 \beta}{(1+z_L)^2} \,{\rm SM} \,
\frac{\Gamma \left(-\frac{\beta}{2} \right)}{\Gamma \left( \frac{\beta}{2} \right)}  \right]^{1/(2-\beta)},  & r_{\rm diff} \gg l_0.  \\
\end{cases}
\end{eqnarray} \label{rdiffDefn}
\end{subequations} 
In the thin-screen approximation one assumes $C_N^2(l)$ to be nonzero across the depth, $\Delta L$, of the phase screen and zero elsewhere.  The quantity ${\rm SM}= \int_l^{l+\Delta L} C_N^2 dl$ is identified as the scattering measure\footnote{Throughout this text we have opted to express the scattering measure in the units of m$^{-17/3}$ rather than the conventional but more cumbersome units of kpc\,m$^{-20/3}$, which are more appropriate in the context of scattering within the Galaxy.}.  This means of calculating the SM is often appropriate for calculating the scattering properties of turbulence inside our Galaxy and, by extension, turbulence in other graviationally collapsed objects, such as the host galaxy of a transient.  
However, the foregoing definition of the scattering measure is inappropriate for objects that are part of the Hubble flow, where the Friedmann-Lema\^itre-Robertson-Walker metric applies.  For a scattering medium that is extended along the line of sight, but in which the thin-screen formalism (i.e. eq.\,(\ref{uGrav})) still applies in an approximate sense (see the discussion in Codona et al.\,1986), the redshift of the scattering, $z_L$, changes continuously, and eqns.\,(\ref{rdiffDefn}) need to be generalized to take this into account.  Furthermore, the integral over path length $\Delta L$ needs to be specified in such a way that it accounts for the geometry of the Universe.  In defining the effective scattering measure, ${\rm SM}_{\rm eff}$ for a medium that extends between redshifts $z$ and $z + \Delta z$, it is useful to absorb the denominator $(1+z_L)^2$ into the definition of the scattering measure and refer all quantities back to the observer's frame so that,
\begin{eqnarray}
{\rm SM}_{\rm eff} &=& \int \frac{C_N^2 (l)}{(1+z')^2} dl = \int_z^{z+\Delta z} \frac{C_N^2(z') d_H(z')}{(1+z')^3} dz',
 \label{SM0eq}
\end{eqnarray}
where we measure distances in terms of light travel time so that $dl = c dt = -d_H(z) dz/(1+z)$ and where, 
\begin{eqnarray}
d_H(z)= c H_0^{-1} [\Omega_\Lambda + \Omega_m (1+z)^3]^{-1/2}, \label{dHubble}
\end{eqnarray} 
is the Hubble radius for a $\Omega=1$ Universe.  In this manner one can still use eqns.\,(\ref{rdiffDefn}) provided one makes the replacement ${\rm SM} / (1+z_L)^2 \rightarrow {\rm SM}_{\rm eff}$. 

For the purposes of numerically computing the magnitude of angular and temporal broadening, we provide numerical expressions for $r_{\rm diff}$ for $\beta=11/3$:
\begin{subequations}
\begin{eqnarray}
r_{\rm diff} &=& \begin{cases}
8.0 \times 10^9 
	\left( \frac{\lambda_0}{1\,{\rm m}} \right)^{-1} 
	\left( \frac{{\rm SM}_{\rm eff}}{10^{12}\,{\rm m}^{-17/3}} \right)^{-1/2}
	 \left( \frac{l_0}{1\,{\rm AU}} \right)^{1/6}  {\rm m}, & r_{\rm diff} < l_0, \\
3.7 \times 10^9 
	\left( \frac{\lambda_0}{1\,{\rm m}} \right)^{-6/5} 
	\left( \frac{{\rm SM}_{\rm eff}}{10^{12}\,{\rm m}^{-17/3}} \right)^{-3/5} {\rm m} , & r_{\rm diff} > l_0. \\
\end{cases}
\end{eqnarray} 
\end{subequations}

\subsection{Angular broadening}

The effect of angular broadening due to plasma turbulence is deduced from the average visibility of the scattered radiation, $\langle V({\bf r}) \rangle = \langle u({\bf X}' +{\bf r}) u^*({\bf X}') \rangle$, where the angular brackets denote an average over the ensemble of phase fluctuations.  Using eq.(\ref{uGrav}) and averaging over the phase fluctuations, the average visibility for a point source of flux density $I_0$ is (Fant\'e 1975; Macquart 2004),
\begin{eqnarray}
\langle V({\bf r}) \rangle &=& \frac{I_0}{(2 \pi r_{\rm F}^2)^2} \int d^2 {\bf x} \,  d^2{\bf x}' 
\exp \left[ 
\frac{i}{2 r_{\rm F}^2} \left( {\bf x} - \frac{D_{LS}}{D_S} ({\bf X}'  + {\bf r}) \right)^2 
- \frac{i}{2 r_{\rm F}^2} \left( {\bf x}' - \frac{D_{LS}}{D_S} {\bf X}' \right)^2 \right]  \nonumber \\
&\null& \qquad \qquad \qquad \qquad \times 
\exp \langle \left[ i \phi({\bf x}) - i \phi ({\bf x}') \right]  \rangle.
\end{eqnarray}
Using the result $\langle \exp [-\phi] \rangle = \exp [-\langle \phi \rangle^2/2]$, and making the change of variable ${\bf R} = {\bf x} - {\bf x}'$ and ${\bf s} = ({\bf x} + {\bf x}')/2$ the average visibility reduces to,
\begin{eqnarray}
\langle V({\bf r}) \rangle =  I_0  \exp \left[ -\frac{1}{2} D_\phi \left( \frac{D_{LS}}{D_S} {\bf r} \right) \right]. \label{Vigm}
\end{eqnarray}
The visibility is related to the image brightness distribution via a Fourier transform, from which we deduce that the angularly broadened image has a radius (half-width at half maximum),
\begin{eqnarray}
\theta_{\rm scat} = f \frac{D_{\rm LS}}{D_{\rm S} \, k \, r_{\rm diff} }, \label{thetaApprox}
\end{eqnarray}
where $f$ is a constant of order unity and $k=2 \pi/\lambda_0$ is the wavenumber in the observer's frame.  (One has $f = 1.18$ if $r_{\rm diff} < l_0$ or $\beta=4$ and $f=1.01$ for $\beta=11/3$ and $r_{\rm diff} > l_0$.)  
A source intrinsically smaller than $\theta_{\rm scat}$ is scatter-broadened to this angular size, whereas the angular sizes of sources intrinsically larger than $\theta_{\rm scat}$ are largely unaltered by scatter broadening.

\subsection{Temporal smearing due to multipath propagation}
Multipath propagation of radiation through a turbulent plasma also causes the signal to be temporally smeared. 
Since the Fresnel-Kirchoff integral in eq.\,(\ref{uGrav}) retains the same form as in Euclidean space, it follows that the form of the expression for the temporal smearing time is identical to that derived in the Euclidean spacetime.  Thus the solution for the smearing timescale proceeds analogously to the Euclidean solution presented in Goodman \& Narayan (1989; GN89).  Specifically, since eq.(\ref{uGrav}) can be cast in an identical form to GN89's eq.\,(2.1.3), the solution for the decorrelation bandwidth proceeds according to the treatment outlined in \S3.3 of GN89.
Thus we find that the scattering timescale associated with a thin screen of scattering material at an angular diameter distance $D_L$ is\footnote{Note that a rigorous derivation of eq.\,(\ref{tauApprox}) is also presented in \S4 of Macquart (2004) where the result is discussed in the context of the gravitational lensing.  The solution is equally applicable to intergalactic scattering since both the lensing and plasma scattering formalisms are founded upon identical propagation equations, as discussed in the context of that work.},
\begin{eqnarray}
\tau = \frac{1}{c k} \left( \frac{r_{\rm F}}{r_{\rm diff}} \right)^2 = \frac{D_L D_{LS} \lambda_0}{2 \pi \, c \, k \, D_S (1+z_L) r_{\rm diff}^2}.  \label{tauApprox}
\end{eqnarray}
An impulsive signal of duration shorter than $\tau$ is smeared by multipath propagation to a timescale of duration $\tau$.
  
Combining relations (\ref{thetaApprox}) and (\ref{tauApprox}) links the angular size of a scattered source to the temporal smearing timescale:
\begin{eqnarray}
\tau = \frac{D_{\rm L} \, D_{\rm S} \, \theta_{\rm scat}^2}{c \, D_{\rm LS} \,(1+z_L)}. \label{tauUse}
\end{eqnarray}
This demonstrates that observations of the angular sizes of radio sources at high redshift either determine or place upper limits on the temporal smearing timescale depending on whether the observed source size represents the scatter-broadening size (when $\theta_{\rm src} < \theta_{\rm scat}$) or the intrinsic source size (when $\theta_{\rm src} > \theta_{\rm scat}$).   The temporal smearing timescale for a given $\theta_{\rm scat}$ depends additionally on the location of the scattering plasma along the line of sight to the source which is, a priori, unknown.  

Numerically, one may express the temporal smearing timescale in terms of the ratio of angular diameter distances $D_{\rm eff} = D_L D_{LS}/D_S$ and the parameters of the turbulence:
\begin{subequations}
\begin{eqnarray}
\tau &=& 
4.1 \times 10^{-5} \, (1+z_L)^{-1} \left( \frac{\lambda_0}{1\,{\rm m}} \right)^4 \left( \frac{D_{\rm eff}}{1\,{\rm Gpc} }\right) \left( \frac{{\rm SM}_{\rm eff}}{10^{12}\,{\rm m}^{-17/3} } \right) \left( \frac{l_0}{1\,{\rm AU}} \right)^{1/3} {\rm s}, \,\, r_{\rm diff} < l_0, \nonumber \\ \label{tauEmpirical1} \\
\tau &=& 1.9 \times 10^{-4} \,(1+z_L)^{-1} \left( \frac{\lambda_0}{1\,{\rm m}} \right)^{22/5} \left( \frac{D_{\rm eff}}{1\,{\rm Gpc} }\right) \left( \frac{{\rm SM}_{\rm eff}}{10^{12}\,{\rm m}^{-17/3} } \right)^{6/5}  {\rm s}, \,\, r_{\rm diff} > l_0. \label{tauEmpirical2}
\end{eqnarray} \label{tauEmpirical}
\end{subequations}


\section{The Scattering Measure of the Intergalactic Medium}

One may regard the theory of angular broadening and temporal smearing at cosmological distances presented in the previous section primarily as a means to reverse-engineer the structure of the turbulent IGM.   However, it is instructive to consider the relative contribution that various scattering regions embedded in the IGM may make to the overall amplitude of the scattering.  It is also illuminating to compare these estimates to the amplitude of the scattering implied by the temporal smearing observed in the six Lorimer bursts detected to date.

There are four obvious components that contribute to the overall scattering measure: (iv) the diffuse Intergalactic Medium, (ii) the turbulent plasma associated with intervening galaxies, (iii) plasma associated with intervening galaxy clusters, and (iv) intervening Lyman-$\alpha$ systems.  We consider contributions from each of these in turn below in terms of their differential contribution to the scattering measure as function of redshift.  To be explicit, we consider the contribution each makes to the quantity $C_N^2(z)$ (i.e. the differential contribution to the scattering measure as a function of redshift)\footnote{Note that in this paper the quantity $z$ is always interpreted as redshift, rather than distance, along the line of sight.}.

\subsection{The diffuse ionized Intergalactic Medium}
Here we consider the contribution that the diffuse ionized IGM makes to the scattering measure.  Since the mean baryonic density of the IGM is well known (Hinshaw et al.\,2013), it is possible to make a relatively simple model that describes the effect of the diffuse IGM on the scattering measure.  

The variance of the electron density fluctuations depends on the amplitude of $C_N^2$ and the outer scale of the turbulent medium $L_0 = 2 \pi q_{\rm min}^{-1}$.  For an electron density power spectrum of the form given in eq.(\ref{PhiDefn}),
the variance in the electron density is, 
\begin{eqnarray}
\langle \delta n_e^2 (z) \rangle 
 &=& 
 C_N^2 \int_{q_{\rm min}=2 \pi/L_0}^{\infty} q^{-\beta} d^3{\bf q}  \\
&\approx& \frac{2 (2\pi)^{4-\beta}}{\beta - 3} C_N^2 L_0^{\beta-3}, \quad L_0 \gg l_0.
\label{CN2DeltaNe}
\end{eqnarray}
The root mean square electron density therefore depends only weakly ($\propto L_0^{1/3}$ for Kolmogorov turbulence) on the outer scale.  However, this parameter is uncertain by many orders of magnitude.  It is plausibly between 0.001\,pc, a scale typically observed in interstellar turbulence (Armstrong, Rickett \& Spangler 1995), and 0.1\,Mpc, the scale of the AGN jets that deposit energy into the IGM.

One can obtain an initial rough estimate of the amplitude of $C_N^2$ by relating the turbulence to the average free electron density of the IGM.  The mean baryonic density of the Universe is 
\begin{eqnarray}
\rho(z) = \frac{3 H_0^2 \Omega_b (1+z)^3}{8 \pi G m_p} =  2.26 \times 10^{-7} \, (1+z)^3 \,\left( \frac{\Omega_b}{0.04} \right) \, {\rm cm}^{-3}, \label{rhoZ}
\end{eqnarray}
for $H_0 = 71\,$km\,s$^{-1}$\,Mpc$^{-1}$ and if we take $\langle \delta n_e^2 \rangle^{1/2}  \sim f \langle n_e(z) \rangle$ this implies, for a Kolmogorov spectrum of turbulence ($\beta=11/3$),
\begin{eqnarray}
C_N^2 (z) = \frac{\beta-3}{2 (2 \pi)^{4-\beta}} L_0^{3-\beta} f^2 \langle n_e(z) \rangle^2  
= 9.42 \times 10^{-14}\, (1+z)^6\, f^2 \, \left(\frac{\Omega_b}{0.04} \right) \left( \frac{L_0}{1\,{\rm pc}} \right)^{-2/3} \,{\rm m}^{-20/3}. \label{CN2diffuse}
\end{eqnarray}

\begin{figure}[h]
\includegraphics[angle=0,scale=0.7]{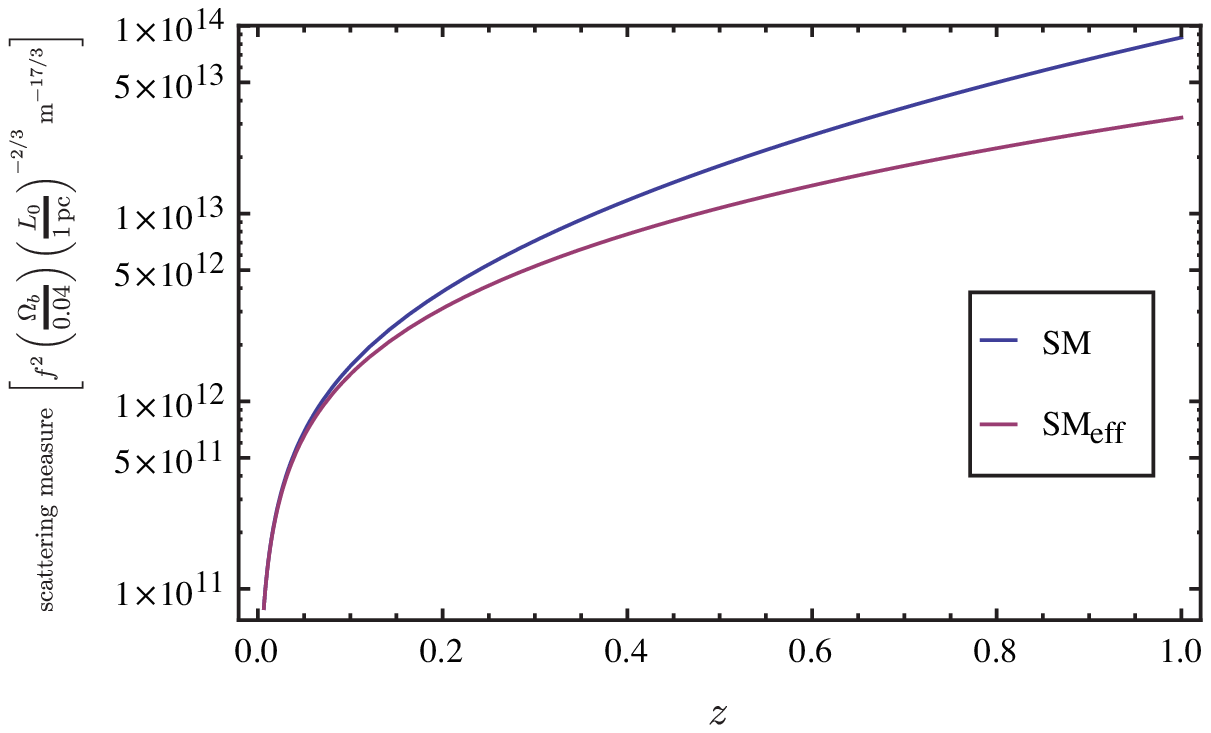}
\includegraphics[angle=0,scale=0.7]{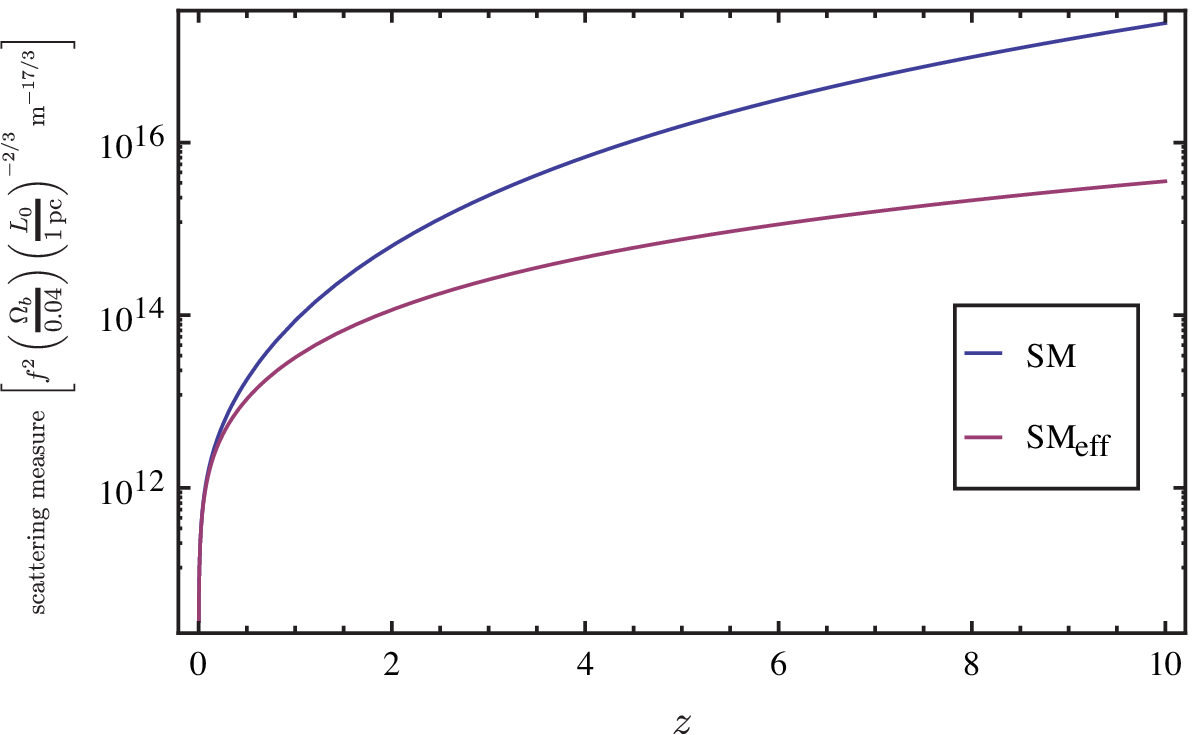}
\caption{Plots of SM and ${\rm SM}_{\rm eff}$ as a function of redshift for a $\Omega_\Lambda=0.7$, 
$\Omega_m=0.3$ Universe.  SM measures the total electron density variance along the line of sight, whereas ${\rm SM}_{\rm eff}$ corrects for the changing wavelength of the radiation as a function of redshift to determine the equivalent phase variance that would be deduced at a redshift of zero on the basis of the density fluctuations along the line of sight.  In scattering theory, the latter is the relevant quantity.} \label{fig:SMplots}
\end{figure}

The scattering measure along the line of sight is then conventionally found by integrating $C_N^2$ along the line of sight,
\begin{eqnarray}
{\rm SM}(z) = \int C_N^2 dl  = \int_0^{z} C_N^2(z') \frac{d_H(z')}{1+z'} dz'.
\end{eqnarray}
For a concordance Universe in which $\Omega=1$, with the radiation contribution $\Omega_r$ assumed negligible, one can use eq.\,(\ref{CN2diffuse}) to predict the scattering measure associated with the diffuse IGM:
\begin{eqnarray}
{\rm SM}(z) &=& 2.73 \times 10^{12} \,
f^2 \, \Omega_m^{-2} \left\{  
3 \Omega_\Lambda - 1 + \left[  1 -3 \Omega_\Lambda + \Omega_m ( (1+z)^3 - 1) \right] \sqrt{(1+z)^3 \Omega_m + \Omega_\Lambda}  
\right\} \nonumber \\
&\null& \qquad \qquad \times \left( \frac{\Omega_b}{0.04} \right) \left( \frac{L_0}{1\,{\rm pc}} \right)^{-2/3} 
\,{\rm m}^{-17/3}.
\end{eqnarray}
While the foregoing definition of the scattering measure is a good indicator of the total electron density variation along the line of sight, it is not useful in evaluating the strength of the scattering because the wavelength of the radiation changes along the ray path and so, therefore, does the amplitude of the phase perturbation for a given density fluctuation.  The correction for this effect is straightforward if the scattering takes place in a narrow range of redshift along the line of sight.  However, the correction is nontrivial if the scattering takes place across a range of redshifts in the IGM.  To take the changing wavelength of the incident radiation into account, it is useful to define an alternate scattering measure, ${\rm SM}_{\rm eff}$ for which the phase perturbations are referred to a standard wavelength (see the discussion following eqns.\,(\ref{rdiffDefn}) above).  In this manner the scattering measure can then be used to determine the cumulative phase variance due to scattering through an extended patch of the IGM.  It is convenient to refer to quantities in the observer frame, for which the wavelength of the observed radiation is denoted $\lambda_0$.  Since the phase perturbation $\delta \phi$ is proportional to $\lambda = \lambda_0 (1+z)^{-1}$, we therefore have ${\rm SM} \propto \delta \phi^2$ and obtain the effective scattering measure of (cf. eq.\,(\ref{SM0eq})),
\begin{eqnarray}
{\rm SM}_{\rm eff}(z) &=& \int_0^{z} \frac{C_N^2(z') d_H(z')}{(1+z')^3} dz'.
\end{eqnarray}
We have evaluated this integral numerically for $\Omega_\Lambda=0.7$ and $\Omega_m=0.3$, and the results are shown in Figure \ref{fig:SMplots}.  We have not found a simple analytic solution for this integral valid over the whole range of $z$.  However, one may expand the integrand to second order in $z$ to find an approximation that is correct to within 10\% at $z<0.7$ or approximate the denominator of the integrand as $(1+z)^{3/2} \Omega_m^{1/2}$ to find an approximation that is correct to within 10\% for $z>2.5$:
\begin{eqnarray}
{\rm SM}_{\rm eff}(z) \approx 
10^{12} f^2 \left( \frac{\Omega_b}{0.04} \right) \left( \frac{L_0}{1\,{\rm pc}} \right)^{-2/3}  \, {\rm m}^{-17/3}
\begin{cases}
3.07  
z \, [ \Omega_m(4 + 3 z) + \Omega_\Lambda (4 + 6z) ]    & z \ll 1, \\
4.91  \, \Omega_m^{-1/2} [(1+z)^{5/2}-1]  & z \gg 1. 
\end{cases} \label{SMeffZ}
\end{eqnarray}
The contribution to the scattering measure for a slice of the intergalactic medium between redshifts $z$ and $z+\Delta z$ is,
\begin{eqnarray}
{\rm SM}_{\rm eff}(z) \approx 
10^{12} \! f^2 \!\! \left( \frac{\Omega_b}{0.04} \right) \!\! \left( \frac{L_0}{1\,{\rm pc}} \right)^{-2/3} \!\!\!\!\! {\rm m}^{-17/3}
\begin{cases}
3.07  
\Delta z \, [\Omega_m (4 + 3 \Delta z + 6 z) + 2 \Omega_\Lambda (2 + 3 \Delta z + 6 z)]     & z \ll 1, \\
4.91  \, \Omega_m^{-1/2} [(1+z+\Delta z)^{5/2}-(1+z)^{5/2}]  & z \gg 1. 
\end{cases}
\end{eqnarray}

\subsubsection{Scattering from a clumpy IGM}

We have considered above the contribution of a diffuse, homogeneously distributed IGM.  However,  scattering measurements from the turbulence in our own Galaxy suggests that this is likely to be too simple a model to adequately capture the full breadth of scattering phenomena likely to be present in the IGM.  Both pulsars and intra-day variable quasars reveal that the turbulent interstellar medium in our own Galaxy is highly inhomogeneously distributed (Cordes \& Lazio 2002) and intermittent (Kowal, Lazarian \& Beresnyak 2007 and Falgarone et al. 2006 and references therein).  There is increasing evidence that many lines of sight are, in fact, dominated not by the diffuse ISM, but instead by single patches of intense turbulence whose scattering measures exceed the expected value from the diffuse ISM by several orders of magnitude; evidence for this is gleaned from  so-called anomalous scattering regions (Cordes \& Lazio 2001), from the ``secondary arcs'' observed in pulsar secondary spectrum, and in Extreme Scattering Events (Putney \& Stinebring 2005; Fiedler et al. 1987).  The occurence of localized pockets of extreme turbulence appears to be a widespread property of turbulent plasmas; some analogous effects appear to occur in terrestrial environments, manifest in the phenomenon of travelling ionospheric disturbances (He et al. 2004).

The physical origin of anomalously strong scattering regions in the ISM is not understood, rendering it unclear how one should extrapolate its properties from the interstellar environment to intergalactic plasmas.  However, the widespread nature of this phenomenon in the ISM suggests that this phenomenon may be too important to ignore when considering possible models of intergalactic turbulence.  To this end, our objective here is to construct a model which phenomenologically captures the effects of anomalous scattering without recourse to a detailed physical model of its origin.


We therefore consider a model that computes the incidence of large (but possibly rare) overdensities as a function of redshift, and then determine how many such regions are likely to exist along a given line of sight.  We start by positing a simple but generic model for the turbulent density fluctuations that enables us to calculate the frequency of overdense regions.  Our basis for the model is a log normal distribution of density fluctuations, whose long tail permits the existence of large high-density deviations from the mean.  Furthermore, simulations and in situ measurements argue that the distribution of density fluctuations in turbulent media is usefully modelled by a log-normal distribution (e.g. V\'azquez-Semadeni 1994; Hopkins 2013 and references therein).  The probability of encountering a clump with a density greater than $N_e$ is 
\begin{eqnarray}
P(X > N_e) 
&=& 1 - \frac{1}{2} {\rm erfc} \left[ \frac{\ln \mu_0 - \frac{1}{2} \ln(f^2+1) -\ln X}{\sqrt{2 \ln (f^2+1)}} \right],
\end{eqnarray}
and the  expected number of clumps whose density exceeds some threshold, $X$, for such a distribution is (see Appendix \ref{ClumpApp}) given by,
\begin{eqnarray}
N (X>N_e;z)
&=& \frac{c }{L_0 H_0} \int_0^z \frac{  1 - \frac{1}{2} {\rm erfc} \left[ \frac{\ln [\mu_0 (1+z)^3] - \frac{1}{2} \ln(f^2+1) -\ln X}{\sqrt{2 \ln (f^2+1)}} \right] }{(1+z)\sqrt{ \Omega_\Lambda + \Omega_m (1+z)^3}} dz. \label{NumberBlobs},
\end{eqnarray}
where $\mu_0$ is the mean baryon density, the root-mean-square density is a factor $f$ times the mean, and we use the fact that the mean density of the IGM increases as $(1+z)^3$.     

\begin{figure}[h]
\includegraphics[angle=0,scale=0.7]{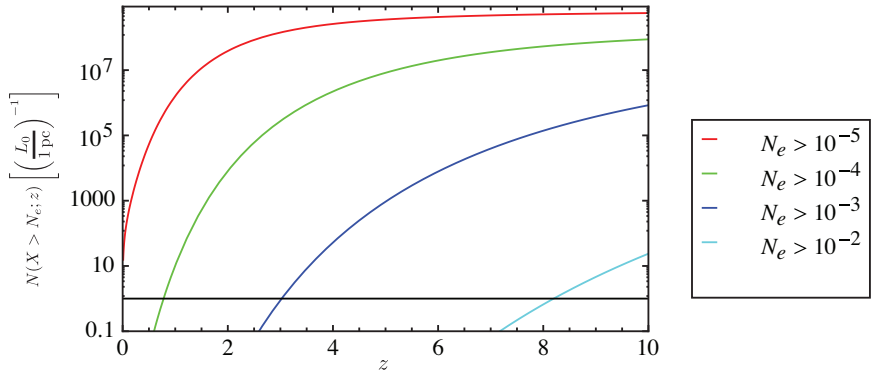}
\caption{The cumulative number of scattering regions exceeding some threshold density $N_e$ along the line of sight as a function of redshift.  The $y$-axis is normalised to a turbulence outer scale length of $L_0 = 1\,$pc; the number of regions encountered is inversely proportional to $L_0$, so that an outer scale of $1\,$Mpc would result in a factor $10^6$ reduction in $N$.  The mean density scales as $(1+z)^3$ as a function of redshift, as per eq.\,(\ref{rhoZ}), and we assume that $f=1$, which is to say that the amplitude of the rms density fluctuation at a given redshift is equal to the mean density at that epoch.  The black horizontal line denotes $N=1$, below which encountering an overdensity of a given $N_e$ is improbable.  } \label{fig:Number}
\end{figure}

To illustrate the behaviour of $N$, we evaluate eq.\,(\ref{NumberBlobs}) for a specific instance.  We set the mean density, $\mu_0$, equal to the mean baryonic density as a function of redshift as per eq.\,(\ref{rhoZ}), and take $f=1$.   Figure \ref{fig:Number} shows the cumulative number of regions whose typical density exceeds various threshold values as a function of redshift.  


Having calculated the number of regions with densities greater than a certain threshold, it remains to compute their contribution to the scattering measure.  We approximate the scattering measure contribution associated with each overdense region as $\sim {C_N^2}_0 L_0= (\beta-3) L_0^{4-\beta} N_e^2/(2 \,  (2 \pi)^{4-\beta}))$.  In this expression for $C_N^2$ we are implicitly treating the outer scale $L_0$ as a proper distance.  This is justifiable in the sense that the outer scale depends only on conditions set by local physics, which is to say that the physics of the turbulence does not evolve with redshift.  It is, of course, possible that even the proper distance $L_0$ does change slowly with redshift if there is some evolution in the overall development of the turbulence throughout the IGM.  However, an investigation of the possible scenarios under which $L_0$ might evolve is beyond the scope of the simple calculations presented here.
Given these provisions, the scattering measure and effective scattering measures are, 
\begin{subequations}
\begin{eqnarray}
{\rm SM} &=& \int_0^z {C_N^2}_0 L_0 \left\lfloor \frac{P(z') d_H(z')}{L_0 (1+z')} \right\rfloor dz', \hbox{ and } \label{SMclumpContribution} \\
{\rm SM}_{\rm eff} &=& \int_0^z \frac{{C_N^2}_0 L_0}{(1+z')^2} \left\lfloor \frac{P(z') d_H(z')}{L_0 (1+z')} 
\right\rfloor dz'. \label{SM0Number}
\end{eqnarray} 
\end{subequations}
We take the integer floor of the quantities inside the symbols $\lfloor \, \rfloor$ because these expressions denote countable quantities (i.e. the number of scattering clouds); the integral picks up an integral amount of scattering, ${C_N^2}_0 L_0$, only each time one encounters a new cloud in the IGM.   

\begin{figure}[h]
\includegraphics[angle=0,scale=0.5]{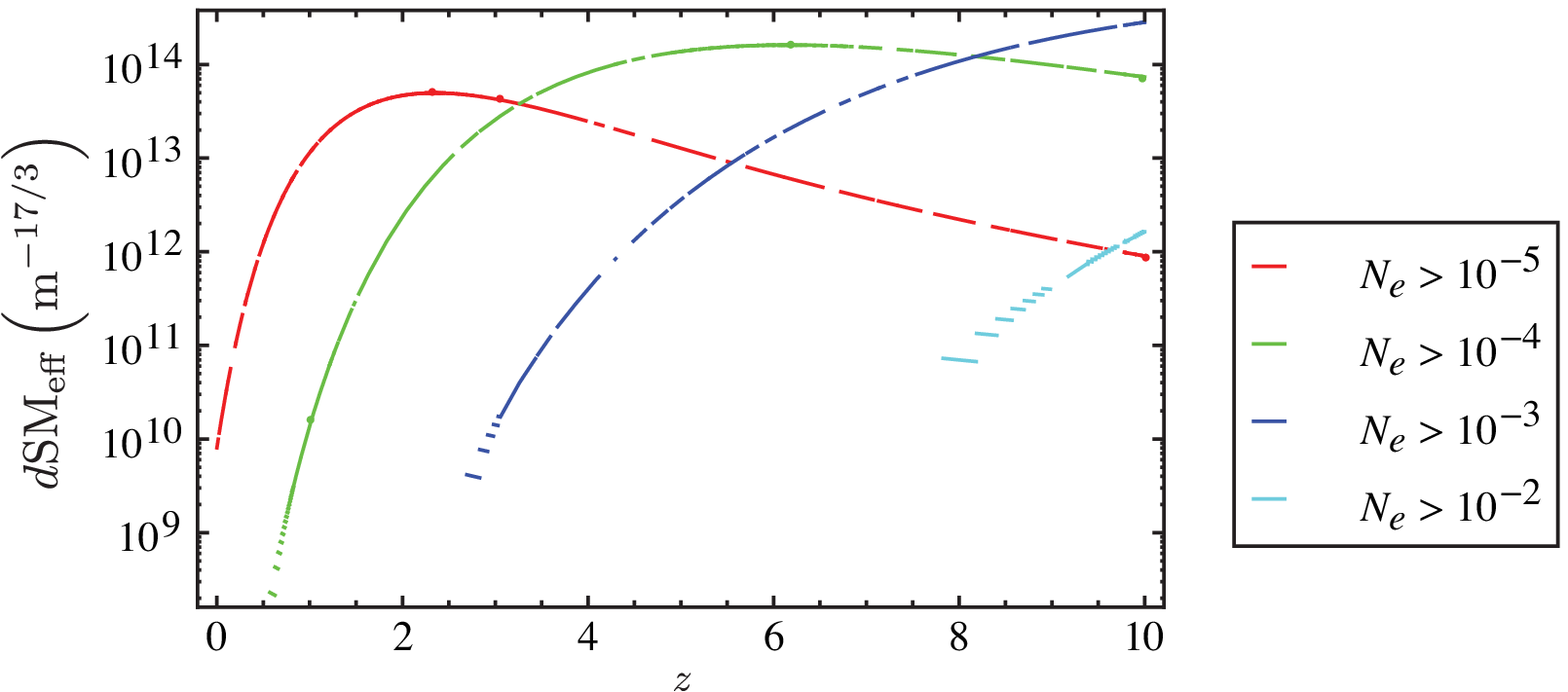}
\includegraphics[angle=0,scale=0.5]{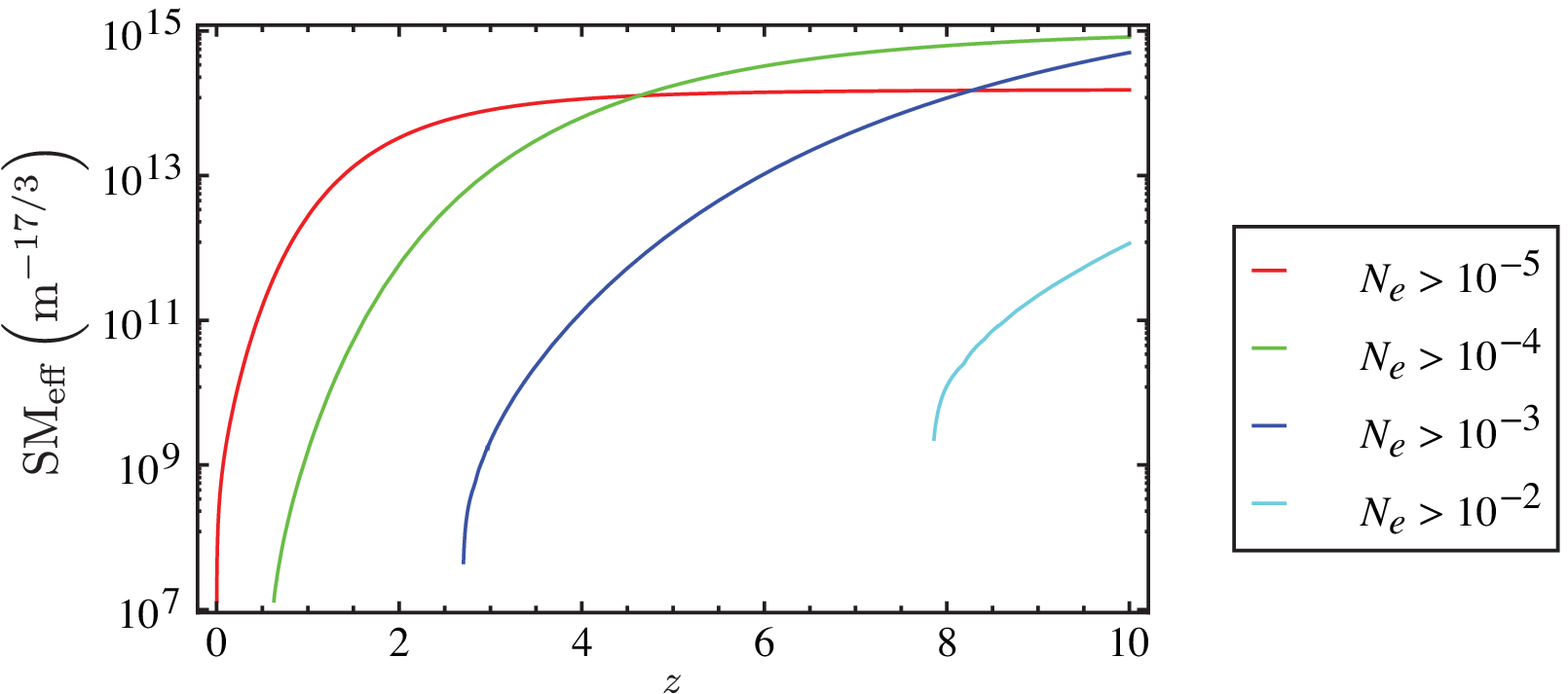}
\caption{Left: the differential contribution to the effective scattering measure as a function of redshift from clouds of various densities in the IGM and Right: the integral effective scattering measure from these same clouds.  Here we assume $f=1$ and that the mean density of the IGM, $\mu_0$, scales as $(1+z)^3$, as per eq.\,(\ref{rhoZ}).} \label{fig:NumberSM0}
\end{figure}

The quantity inside the integrand in eq.\,(\ref{SM0Number}) may be considered an effective incremental element of scattering measure as function of redshift, which we denote $d{\rm SM}'_0$.  We plot the value of both $d{\rm SM}'_{\rm eff}$ and ${\rm SM}_{\rm eff}$ in Figure \ref{fig:NumberSM0}.  The behavior of the curves of $d{\rm SM}_{\rm eff}$ in this Figure may be understood as follows.  At low redshifts only moderate density clouds, $N_e \sim 10^{-5}\,$cm$^{-3}$, are common enough to make a substantial contribution to the scattering measure.  However, as the mean density of the IGM increases as $(1+z)^3$, gradually larger and larger overdensities become more probable.  As the increment of SM associated with denser clouds is larger, there comes a point at which their contribution to $d{\rm SM}_{\rm eff}$ dominates over that from the more numerous but less dense clouds.  Thus, at low redshifts clouds of densities $\sim 10^{-5}\,$cm$^{-3}$ dominate, but at $z \sim 3$ clouds of densities $\sim 10^{-5}\,$cm$^{-3}$ dominate the contribution to $d{\rm SM}_{\rm eff}$, and at $z \sim 8$ clouds of densities $\sim 10^{-3}\,$cm$^{-3}$ begin to dominate the contribution to $d{\rm SM}_{\rm eff}$.  

\subsection{The contribution of galaxies and intracluster gas}

We now consider scattering by intervening gravitationally bound objects, namely the intracluster medium (ICM) of rich clusters of galaxies, as well as the ISM of other galaxies. We assume that the electron densities and turbulent properties of these objects are constant (with no evolution) across the redshifts of interest. The values of $C_N^2$ for these objects are therefore independent of redshift, and are given by:
\begin{equation}\label{CN2gal}
C_{N,{\rm gal}}^2 = 1.8 \times 10^{-3} {\left( \frac{n_e}{10^{-2} \,{\rm cm^{-3}}} \right)}^{2} {\left( \frac{L_{0}}{0.001 \,{\rm pc}} \right)}^{-2/3} {\rm m^{-20/3}}
\end{equation} 
for scattering through a spiral galaxy similar to the Milky Way, and 
\begin{equation}\label{CN2icm}
C_{N,{\rm icm}}^2 = 8.4 \times 10^{-13} {\left( \frac{n_e}{10^{-4} \,{\rm cm^{-3}}} \right)}^{2} {\left( \frac{L_{0}}{0.1 \,{\rm Mpc}} \right)}^{-2/3} {\rm m^{-20/3}}
\end{equation}  
for scattering through the ICM. Kolmogorov turbulence is assumed, and the parameters $L_{0}$ and $n_e$ are normalized by values appropriate for galaxies and the ICM respectively. 

The scattering measure, ${\rm SM}$, is obtained simply by multiplying $C_{N,{\rm obj}}^2$ with the thickness of the object ($\Delta L$) and number of objects ($N_{\rm obj}$) intersecting the line-of-sight to the background source:
\begin{equation}\label{SMicmgal}
{\rm SM} = C_{N,{\rm obj}}^2 \, \Delta L \,N_{\rm obj},
\end{equation}
where $C_{N,{\rm obj}}^2$ can be $C_{N,{\rm gal}}^2$ or $C_{N,{\rm icm}}^2$. Note also that the effective scattering measure, ${\rm SM_{eff}}$, is dependent on the redshifts of the intervening objects, so that:
\begin{equation}\label{SMefficmgal}
{\rm SM_{eff}} = {\rm SM} \sum_{i = 1}^{N_{\rm obj}} {(1+z_{L,i})^{-2}},
\end{equation}
The mean number of objects intersecting a background source at a redshift of $z$ is given by 
(Padmanabhan 2002) as: 
\begin{equation}\label{intersectobject}
N_{\rm obj}(z) = \int_0^z \frac{\pi r^2n(z')d_H(z')}{(1+z')} dz',
\end{equation}  
where $r$ and $n(z)$ are the typical proper radius and proper number density of the object respectively.  

\begin{figure}[h]
\includegraphics[angle=0,scale=0.8]{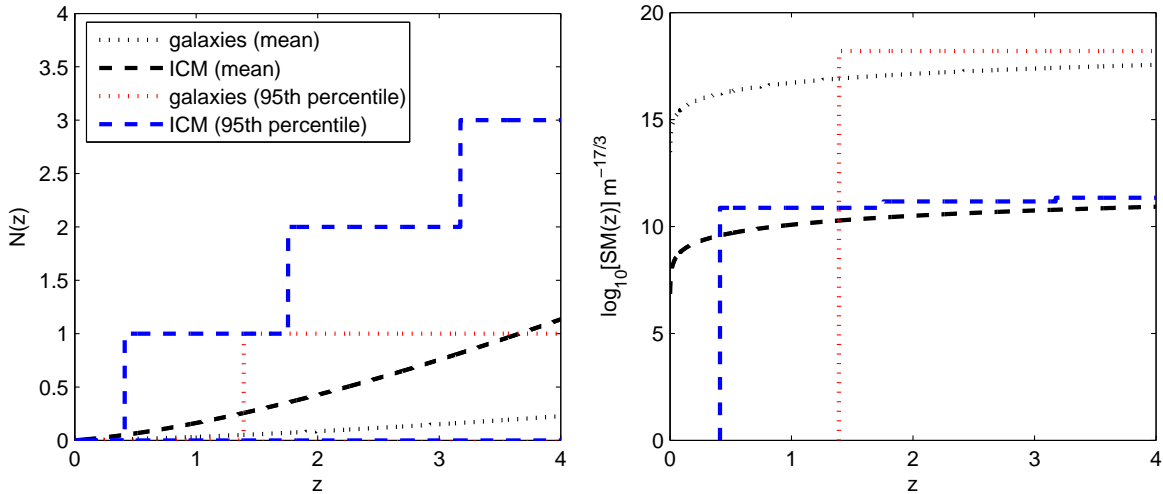}
\caption{The number of galaxies and clusters of galaxies intersecting a source at a redshift of $z$ (left panel), as well as the scattering measure as a function of $z$ (right panel). The black lines give the mean values, while the colored curves show the 95th percentile based on Poisson statistics. We have used the parameters in Table~\ref{galicmparam}.} \label{fig:probgalcluster}
\end{figure}

We evaluate the ${\rm SM}$ for galaxies and the ICM, assuming that their sizes do not evolve, and that their number densities are conserved over the redshifts of interest, so that $n(z) = n_0(1+z)^3$, where $n_0$ is the number density at the present epoch. We used typical values for the various parameters, listed in Table~\ref{galicmparam}. Both the mean $N(z)$ and ${\rm SM(z)}$ for galaxies and the ICM are shown as black curves in Figure~\ref{fig:probgalcluster}. We also evaluate the probability of intersecting $N = 0,1,2...$ objects at each redshift, taking the statistics of object counts to follow a Poisson distribution.
 
\begin{deluxetable}{l c c c}

\tabletypesize{\footnotesize}
\tablewidth{0pt}
\tablecaption{Parameters used for galaxies and the ICM. \label{galicmparam}}
\tablehead{
\colhead{Parameter} & \colhead{Symbol} & \colhead{Galaxy} & \colhead{ICM}
}

\startdata
mean electron density  & $n_e$ & $10^{-2} \,{\rm cm^{-3}}$ & $10^{-4} \,{\rm cm^{-3}}$\\
outer scale of turbulence & $L_{0}$  & 0.001 pc & 0.1 Mpc\\
proper radius & $r$ & $10h^{-1}$ kpc & $1 h^{-1}$ Mpc \\
proper number density & $n_0$ & $0.02h^3\, {\rm Mpc}^{-3}$ & $10^{-5}h^3\, {\rm Mpc}^{-3}$\\

\enddata


\end{deluxetable}

The value of $C_N^2$ for individual galaxies is relatively large, typically $\sim 1.8 \times 10^{-3}\, {\rm m^{-20/3}}$, thus their ${\rm SM}$ is at least $\sim 2$ orders of magnitude  larger (depending on source redshift) than that predicted for the diffuse IGM. However, due to their scarcity and smaller sizes, their contribution is unlikely to dominate for most sight-lines, since the mean number of intersecting galaxies is $\ll 1$ up to $z \sim 4$, and particularly at $z \lesssim 1.5$ where there is a $95\%$ chance that the source is not intersected by a galaxy, following Poisson statistics. In the event that a transient source is indeed intersected by an intervening galaxy, the ${\rm SM_{eff}}$ would then depend on where the scattering galaxy is located along the line-of-sight, $z_{\rm gal}$, and thus be reduced by a factor $(1 + z_{\rm gal})^2$. The ISM of such intervening galaxies, especially those located at $z \sim 0$, would then dominate the SM towards the transient source.   

While the chances of intersecting a galaxy cluster is higher due to its larger size, approaching an average of 1 intervening cluster for a source at $z \sim 4$, the typical value of $C_{N,{\rm icm}}^2$ is much lower, $\sim 8.4 \times 10^{-13}\, {\rm m^{-20/3}}$, due to the lower electron densities and the larger outer scales of turbulence used in the approximations. The ICM therefore also does not contribute much to the overall SM on average, unless the $L_{0}$ of the ICM is of order $\sim 1$pc, at which the contribution to the SM will be comparable to that shown in Figure~\ref{fig:NumberSM0}.

We thus conclude that the ISM of intervening galaxies and the ICM are important if they do intersect sight-lines towards transient sources, but do not contribute significantly to the total ${\rm SM}$ for most sight-lines. The statistics used, however, do not account for the fact that extragalactic transient sources are more likely to be detected within a large cluster of galaxies, or that searches for extragalactic transients will most likely be directed towards galaxies or galaxy clusters.

\subsection{The contribution of Lyman-$\alpha$ systems}

The ionized material associated with Lyman-$\alpha$ systems potentially contributes to the scattering of extragalactic sources.  The magnitude of this contribution was investigated by Rickett et al.\,(2007), whose results we briefly summarize for completeness here. They used the Haardt-Madau cosmological model for reionization by the UV background (Haardt \& Madau 1996), and assume reionization equilibrium, to infer the electron densities of photoionized gas associated with Lyman-Alpha clouds, whose HI column densities are well studied over a wide range of redshifts (Prochaska et al.\,2005; Janknecht et al. 2006). As with our models, these clouds are assumed to have fully developed turbulence with a Kolmogorov spectrum. Applying the thin screen scattering model, and summing over the scattering contributions of all clouds in the line of sight, they predict angular broadening of order $\sim 1 \,\mu$as up to $z \sim 5$ at 5 GHz, for cloud sizes of order $\sim 10$ kpc. Based on the relation in Equation~\ref{tauUse}, this translates to temporal smearing timescales of order $\sim 1$ to $10 \mu$s at 5 GHz, depending on the locations of the scattering clouds, and $\sim 1$ ms to $10$ ms at 1 GHz, which are comparable to that of the other components already discussed, as well as the limits imposed by the Lorimer bursts, which we discuss below. 

\subsection{An empirical estimate of the DM-SM relation} \label{DMSM}

To supplement the foregoing estimates, we also consider an alternate empirical approach based on attempts to relate dispersion measure and scattering measure.  An empirical relation between DM and SM is observed in turbulent astrophysical plasmas in the interstellar medium (Bhat et al. 2004), and it is therefore useful to consider how this relationship might relate to turbulence in intergalactic plasmas.   This approach holds practical appeal because the dispersion measure is a fundamental measureable of any implusive cosmological transient, and a relation between SM and DM would suggest a characteristic DM out to which scattering may begin to assert its importance in limiting the detectability of extragalactic transients.

Following the formalism of Cordes et al. (1991) and Bhat et al. (2004), one may consider the IGM to be comprised of a collection of scattering clouds in which the respective DM  and SM increments associated with a scattering cloud of depth $\delta s$ are
\begin{eqnarray}
\delta {\rm DM} &=& n_e \delta s \quad \hbox{ and } \\
\delta {\rm SM} &=& C_{\rm SM} F_c n_e^2 \delta s,
\end{eqnarray}
where $C_{\rm SM}$ takes the value 1.84\,m$^{-20/3}$cm$^6$ if we assume a Kolmogorov spectrum of turbulence, and where $\delta s$ is expressed in kpc, $n_e$ in cm$^{-3}$, DM in kpc\,cm$^{-3}$ and SM is expressed in its usual units of kpc\,m$^{-20/3}$.  The dimensionless fluctuation parameter is defined by 
\begin{eqnarray}
F_c = \frac{\zeta \epsilon^2}{\eta} \left( \frac{L_0}{1\,{\rm pc}} \right)^{2/3} \label{FcDefn}
\end{eqnarray}
where $L_0$ is the outer scale of the turbulence, $\eta$ is the volume filling factor of the turbulent medium, and $\epsilon = \langle (n_e - \bar{n_e})^2 \rangle/ \langle n_e \rangle^2$ measures the amplitude of density fluctuations within the cloud relative to its mean value, and $\zeta = \langle \overline{n_e^2} \rangle /\langle \overline{n_e} \rangle^2$ is a measure of fluctuations in the mean density between clouds.

This implies the following relation between increments in DM and SM,
\begin{eqnarray}
\frac{\delta {\rm SM}}{\delta {\rm DM}} = 1.84 \langle n_e \rangle F_c  
\end{eqnarray}
where the fluctuation parameter is found to lie in the range $0.20 - 110$ for turbulence in the interstellar medium of our Galaxy (Cordes \& Lazio 2002 (NE2001)).


The extension of this formalism to an intergalactic context involves two modifications which alter the ratio of the SM to the DM.  
1. The mean density of the IGM ($2.3 \times 10^{-7} (1+z)^3$\,cm$^{-3}$) is much lower relative to the ISM ($0.02\,$cm$^{-3}$), so that a given dispersion measure variation is expected to produce an SM contribution that is a fraction $10^{-5} (1+z)^3$ of that expected in the ISM.  2. The characteristic cloud thickness and outer turbulent scale may be much larger in the IGM than in the ISM.  If we associate with outer scale of intergalactic turbulence with the scale of sources likely to inject turbulent power, such as AGN jets, it may be more plausible to identify the outer scale for intergalactic turbulence with scales $L_0 \sim 0.1\,$Mpc, in which case the fluctuation parameter $F_c$ would be a fraction $5 \times 10^{-4}$ that of the ISM.  However, we caution that the effective outer scale may be lower by many orders of magnitude, as it may instead be dictated by microphysical properties of the turbulence (i.e. it may be as low as $1\,$pc), in which case one might expect values similar to those encountered in the ISM .  Given these considerations, one therefore expects $\delta {\rm SM}/\delta {\rm DM} \sim 4.2 \times 10^{-7} (1+z)^3 F_c$.  With $\delta {\rm DM}$ expressed in the more conventional units of pc\,cm$^{-3}$ and SM in units of m$^{-17/3}$, one finds
\begin{eqnarray}
\delta {\rm SM} \sim 1.3 \times 10^{13} \, (1+z)^3 \, F_c  \left( \frac{\delta {\rm DM}}{1000\,{\rm pc\,cm}^{-3}} \right)  \, {\rm m}^{-17/3}.
\end{eqnarray}

While this approach presents only a crude means of estimating the relationship between DM and SM in the intergalactic medium, due to the inherent uncertainty in the estimate of $F_c$ in a plasma whose turbulent properties may qualitatively differ from the ISM, it does demonstrate two important properties of the turbulence. Firstly, the nature of the $\delta$DM-$\delta$SM relationship is not constant, but is expected to increase sharply with redshift, scaling as $(1+z)^3$.  Secondly, it demonstrates that the amount of scattering expected per unit of dispersion measure is likely to be several orders of magnitude lower than that found in the ISM.  The range of values implied by this approximation is consistent with the order of magnitude of the SM computed by other means in previous subsections.  
The lower amplitude of scattering per unit DM suggests that level of temporal smearing experienced by extragalactic bursts is substantially lower than may be expected on the basis of extrapolation from the DM-SM relationship observed in the ISM, and there thus are excellent prospects of detecting extragalactic bursts out to high redshifts.

The SM estimate based on the detection of temporal smearing in FRB 110220, as discussed above, may be used to provide a tentative estimate of $F_c$ in the IGM.  For this burst, we use $z=0.81$ (Thornton et al. 2013) to find $F_c = 1.3 \times 10^2$.  This value is similar to that obtained in highly turbulent regions of the ISM, and it suggests that the outer turbulent scale may not be dissimilar from that found in interstellar environments.


\section{Existing scattering limits and the redshift dependence of temporal smearing} 

\subsection{Comparison with existing angular broadening limits}

It is worth considering whether high resolution VLBI images of high redshift radio sources pose any additional constraints on the angular broadening, and hence temporal smearing.  In terms of the quantities derived in this paper, the angular size of the source is
\begin{subequations}
\begin{eqnarray}
\theta &=& \begin{cases} 4.8 \left( \frac{\lambda}{1\,{\rm m}} \right)^2 \left( \frac{D_{LS}}{D_S} \right) \left( \frac{{\rm SM}_{\rm eff}}{10^{12}\,{\rm m}^{-17/3}} \right)^{1/2} \left( \frac{l_0}{1\,{\rm AU}} \right)^{-1/6} \mu{\rm as}, & r_{\rm diff} < l_0, \\
8.9 \left( \frac{\lambda}{1\,{\rm m}} \right)^{11/5} \left( \frac{D_{LS}}{D_S} \right) \left( \frac{{\rm SM}_{\rm eff}}{10^{12}\,{\rm m}^{-17/3}} \right)^{3/5}  \mu{\rm as},  & r_{\rm diff} > l_0. \\
\end{cases}
\end{eqnarray}
\end{subequations}
The largest values of ${\rm SM}_{\rm eff}$ indicated by our various models are of the order of $3 \times 10^{13}\,$m$^{-17/3}$ for $z<1$ and $10^{15} \,$m$^{-17/3}$ for $z<5$, indicating that a conservative upper bound on the expected scattering broadening size is $\sim 100 \lambda^2 \,\mu$as for $z \la 1$ and $\sim 600 \lambda^2 \,\mu$as for $z \la 5$. 

The most stringent observational limits come from VLBI and from the minimum source angular sizes of sources as deduced by interstellar scintillation at cm-wavelengths.  The strongest VLBI limits are at the lowest frequencies since the expected angular broadening size increases quadratically with wavelength but the resolution of an interferometer degrades only linearly with wavelength.   However, the limit on IGM angular broadening at $z \la 1$ is well beyond the resolving capabilities of VLBI: baselines larger than $2.0 \times 10^6\,$km are required at 300\,MHz to access sources at the angular resolutions relevant to IGM angular broadening.  At higher frequencies constraints from ISS are more stringent.  

Scintillation studies of intra-day variable sources yield source lower limits on the source size of $\sim 10\,\mu$as (e.g. Rickett, Kedziora-Chudczer \& Jauncey 2002; Macquart \& de Bruyn 2007) at 5\,GHz.  However, the predicted upper limit on angular broadening by the IGM at this frequency is $\sim 0.4\,\mu$as, far lower than the lower limits deduced from ISS.  

We may also utiliize the angular sizes of intra-day variable sources to place upper limits on the effective scattering measure in the IGM at moderate redshifts.  The $z=0.54$ source J1819$+$3845, contains components between 9 and 26$\,\mu$as in size at 5\,GHz (Macquart \& de Bruyn 2007).  The smallest of these components therefore implies ${\rm SM}_{\rm eff} < 9.8 \times 10^{16}\,$m$^{-17/3}$, where we assume $D_{LS}/D_S =0.5$.   Similarly, the smallest component size deduced for the quasar PKS 0405$-$385 is 9\,$\mu$as (Rickett et al. 2002), which places an identical limit on the effective scattering measure at $z=1.285$.


\subsection{Comparison with existing burst characteristics} 
The temporal characteristics of the six extragalactic bursts reported by Lorimer et al.\,(2007), Keane et al.\,(2011) and Thornton et al. (2013) may be used to further constrain the properties of scattering over cosmological distances.  Temporal smearing was detected in only two bursts, and only the detection of the highest S/N event reported by Thornton et al.\,(2013) was not of marginal significance: this burst, FRB 110220, possessed a (smearing-dominated) duration of $5.6 \pm 0.1\,$ms duration at 1.3\,GHz.  No significant detection of scattering was made in other bursts, placing limits on the smearing timescale of between between $<1.1\,$ms and $<4.3$\,ms at the same frequency. Moreover, we note that the upper limit on the DM$=1072\,$pc\,cm$^{-3}$, $<4.3\,$ms duration burst (FRB 110703) is smaller than the duration of FRB 110220, despite occurring at the lower DM of $910\,$pc\,cm$^{-3}$. This indicates that there is appreciable variation in the smearing timescale of extragalactic bursts between different lines of sight.

We use eqns.\,(\ref{tauEmpirical1}-\ref{tauEmpirical2}) to determine the effective scattering measure implied by these durations.
The frequency dependence of the temporal smearing observed in FRB 110220, $\tau \propto \nu^{-4.0 \pm 0.4}$ is consistent with both scattering in which $r_{\rm diff} < l_0$ ($\tau \propto \nu^{-4}$) or in which $r_{\rm diff} > l_0$ ($\tau \propto \nu^{-4.4}$).  Since the observed pulse shape is the convolution of the intrinsic pulse shape with the pulse broadening kernel, we attribute 4\,ms to the temporal smearing timescale in FRB 110220, which implies the following 
\begin{eqnarray}
{\rm SM}_{\rm eff} = (1+z_L) \left( \frac{D_{\rm eff}}{1\,{\rm Gpc}} \right)^{-1} \left\{ \begin{array}{ll} 
3.4 \times 10^{16} \left( \frac{l_0}{1\,{\rm AU}}\right)^{-1/3} \,{\rm m}^{-17/3}, & r_{\rm diff} < l_0, \\
1.8 \times 10^{16}\,{\rm m}^{-17/3}, & r_{\rm diff} > l_0. \\
\end{array} \right.
\end{eqnarray}
This may be regarded as an upper limit on the effective scattering measure of the IGM in the sense that some component of the temporal smearing may originate in the interstellar medium of the host galaxy.
More generally, the absence of broadening in the other bursts implies 
\begin{eqnarray}
{\rm SM}_{\rm eff} < (1+z_L) \left( \frac{\tau}{1\,{\rm ms}} \right) \left( \frac{D_{\rm eff}}{1\,{\rm Gpc}} \right)^{-1} \left\{ \begin{array}{ll} 
8.6 \times 10^{15} \left( \frac{l_0}{1\,{\rm AU}}\right)^{-1/3} \,{\rm m}^{-17/3}, & r_{\rm diff} < l_0, \\
4.7 \times 10^{15}\,{\rm m}^{-17/3}, & r_{\rm diff} > l_0. \\
\end{array} \right.
\end{eqnarray}
The scattering is particularly constraining for FRBs 110627 and 120127, with durations of $<1.4\,$ms and $<1.1\,$ms respectively (Thornton et al.\,2013).

\subsection{The distinction between scattering due to the host galaxy and the IGM}
The extension of the scattering theory above to cosmological distances allows us to immediately address an important issue related to the origin of scattering in high-DM Lorimer-like bursts.    Specifically, in several high-DM bursts the pulse width is observed to increase as $\lambda^{4.0 \pm 0.4}$, which suggests that these pulses are broadened by multipath propagation by an inhomogeneous plasma along the lines of sight to the objects.  For the case of the Lorimer burst itself, the predicted temporal smearing time due to our Galaxy's interstellar medium is $0.6\,\mu$s at 1.4\,GHz (Cordes \& Lazio 2002), which is much smaller than the observed $\approx 5\,$ms pulse duration.  Thus the scattering occurs either in the IGM or in the host galaxy of the burst itself.   There is currently considerable debate about whether the observed temporal smearing is caused by the IGM, or whether it can be fully attributed to turbulence in the interstellar medium of the galaxy in which the burst occured (the host galaxy).  

The foregoing theory makes a strong prediction about the redshift dependence of the scattering at cosmological distances.  This potentially enables us to distinguish between temporal smearing predominately due to the host galaxy or the IGM itself.  If the scattering is dominated by turbulent plasma inside the host galaxy, the SM will reflect local conditions within those galaxies and will be decoupled from the Hubble expansion that affects the density of the diffuse IGM.  
We discuss the effect of temporal smearing in relation to eq.\,(\ref{tauApprox}).  For scattering occurring at the host galaxy one has, to an excellent approximation, $D_L / D_S = 1$ so, 
\begin{eqnarray}
\tau_{\rm host} = \frac{D_{LS} \lambda_0^2}{4 \pi c  \, (1+z_L) \, r_{\rm diff}^2}. \label{tauhost}
\end{eqnarray}
Thus, if one were to suppose that the values of SM of the various host galaxies are comparable between different bursts, since they reflect the instrinic properties of the environments of the host galaxies rather than the evolving density of the IGM with redshift, then one has $r_{\rm diff} \propto (1+z_L)$ and $\tau \propto (1+z_L)^{-3}$ under the assumption that the diffractive scale is smaller than the inner scale of the turbulence\footnote{The temporal smearing timescale scales as $\tau \propto \lambda^4$ if the diffractive scale is smaller than the inner scale of the turbulence.  In the opposite regime, $r_{\rm diff} > l_0$, one has $\tau \propto \lambda^{2 \beta/(\beta-2)}$.  The observed $\lambda^4$ dependence of temporal smearing of the high-DM pulses suggests that $r_{\rm diff} < l_0$.}.  If, instead, the diffractive scale is larger than the inner scale of the turbulence, $r_{\rm diff} > l_0$, one has $\tau \propto (1+z)^{(2+\beta)/(2-\beta)}$.  Thus, for objects in which the scattering is dominated by the interstellar medium of the host galaxy, this leads to the surprising result that the temporal smearing time will {\em decrease} as a steep function of redshift!

This result is in stark contrast to the redshift dependence of the temporal smearing time expected of scattering in the IGM.  Equation (\ref{SMeffZ}) shows that, for $z \ll 1$, the effective scattering measure scales linearly with redshift, while for $z \gg 1$ it scales proportional to $(1+z)^{2.5}$.  Using eq.\,(\ref{tauApprox}), we can derive the redshift dependence of the scattering time in terms of the redshift of the burst, $z_S$:
\begin{eqnarray}
\tau_{\rm IGM} \propto \frac{D_L D_{LS}}{D_S} 
\begin{cases}
z_S & z_S \la 1, \\
(1+z_S)^{3/2} & z_S \ga 1.
\end{cases} \label{tauIGMLowHighZ}
\end{eqnarray}
This result is subject to the assumption that the diffractive scale is smaller than the inner scale of the turbulence; this assumption is supported by the observed frequency dependence of the temporal smearing of the FRBs observed so far.  For completeness, however, we also quote the results for the case in which the diffractive scale exceeds the inner scale of the turbulence:
\begin{eqnarray}
\tau_{\rm IGM} \propto \frac{D_L D_{LS}}{D_S} 
\begin{cases}
z_S^{2/(\beta-2)} & z_S \la 1, \\
(1+z_S)^{(7-\beta)/(\beta-2)} & z_S \ga 1.
\end{cases}
\end{eqnarray}
The redshift dependence of the ratio $\zeta=D_L D_{LS}/D_S$ must also be taken into account when examining the temporal smearing time.   However, at $z_S \ga 1$ this ratio generally exhibits only a weak dependence on source redshift, and therefore does not strongly alter the redshift dependence of the temporal smearing timescale.  

\begin{figure}[h]
\centerline{\includegraphics[angle=0,scale=1.0]{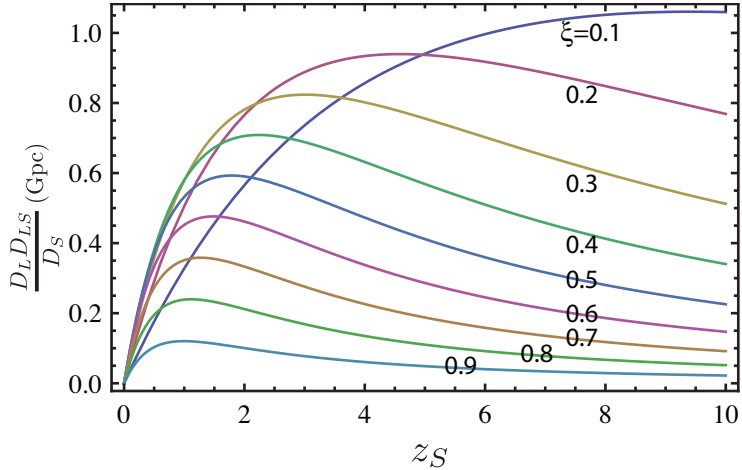}}
\caption{The ratio $\xi=D_L D_{LS}/D_S$ as a function of the source redshift, $z_S$, and the effective redshift of the scattering material, $z_L= \xi z_S$.} \label{fig:DLratio}
\end{figure}

Figure \ref{fig:DLratio} shows the evolution of the ratio $\zeta$ as a function of the source redshift, $z_S$ and the effective location of the scattering material, whose redshift we parameterise in terms of the fraction $\xi$ of the source redshift, $z_L = \xi z_S$.  It is evident that, for most of the range of effective screen distances (i.e. $0.3 \la \xi \la 0.9$), the ratio $\zeta$ turns over and is nearly flat in the range $1 \la z_S \la 2$, and then declines slowly to higher redshifts.  For $z \ga 3$ the decline in $\zeta$ with source redshift scales as $(1+z_S)^{-\alpha}$, with the index $\alpha$ increasing from $\approx 0.5$ at $\xi=0.3$ to a value of $\alpha \approx 1.3$ for $\xi=0.9$.  In the regime of very small source redshifts, $z_S \ll 1$, in which redshift is approximately linearly proportional to the angular diameter distance, we obtain the simple result $\zeta \approx c H_0^{-1} z_L (z_S - z_L)/[z_S (1+z_L)] \approx z_L(1 -z_L/z_S)$.

Combining the redshift scaling of $\zeta$ with the scaling expected of IGM turbulence in eq.\,(\ref{tauIGMLowHighZ}), we obtain an overall redshift dependence of,
\begin{eqnarray}
\tau_{\rm IGM} \propto  
\begin{cases}
z_S^2 & z_S \la 1, \\
(1+z_S)^{0.2-0.5} & z_S \ga 1.
\end{cases} \label{tauIGMFinal}
\end{eqnarray}
Thus we see that at low redshifts the temporal smearing time is expected to increase quadratically with redshift, but turns over and only increases as weakly with source redshift beyond $z_S \sim 1$.  

The foregoing results present a definitive statistical means of determining the origin of the scattering observed in high-DM bursts.  There is a strong difference in the redshift dependence of scattering due to the interstellar medium of the burst's host galaxy or whether it is due to the IGM.  The scattering timescale of bursts that are scattered predominately by the host galaxy scales as $(1+z_{\rm host})^{-3}$, whereas the timescale of bursts scattered by the IGM increase as $z_S^2$ for low redshifts, and in the range $(1+z_S)^{0.2}$ to $(1+z)^{0.5}$ for bursts at $z_S \ga 1$.

It should be emphasised that this distinction applies only in a statistical sense, in that there is likely to be considerable variation between the scattering properties of bursts.  For instance, one expects some variation in the SM of the host galaxy depending on its orientation relative to the line of sight.  The scattering of the host galaxy is more likely to dominate over the IGM contribution on occasions when the burst is viewed edge-on through the host galaxy.  However, the fraction of bursts in which orientation plays a strong role is likely to be small.  Experience in our own Galaxy reveals that interstellar scattering decreases sharply at Galactic latitudes $|b| \ga 15^\circ$ so that, in a galaxy with scattering properties similar to the Milky Way, only $3$\% of bursts would be viewed the a galactic disk sufficiently inclined for it to contribute substantially to the scattering.


%


\section{Conclusions}

Our conclusions are summarised as follows:
\begin{itemize}
\item We have generalized the theory of scattering to take into account the curved geometry of spacetime at cosmological distances.  In curved spacetime the temporal smearing timescale can be expressed in the form
\begin{eqnarray}
\tau  = \frac{D_{\rm eff} \lambda}{2 \pi c k (1+z) r_{\rm diff}^2},
\end{eqnarray}
where $D_{\rm eff} = D_L D_{LS}/D_S$ is the effective scattering distance expressed in terms of angular diameter distances between the observer and the medium ($D_L$), the observer and the source ($D_S$) and the medium and the source ($D_{LS}$), and $r_{\rm diff}$ is the diffractive scale of the scattering medium, which is a function of the scattering measure.  Numerically, one has
\begin{eqnarray}
\tau = 
(1+z_L)^{-1} \left( \frac{D_{\rm eff}}{1\,{\rm Gpc} }\right) \left( \frac{{\rm SM}_{\rm eff}}{10^{12}\,{\rm m}^{-17/3} } \right)
\left\{ \begin{array}{ll}
4.1 \times 10^{-5} \,  \left( \frac{\lambda_0}{1\,{\rm m}} \right)^4  \left( \frac{l_0}{1\,{\rm AU}} \right)^{1/3} {\rm s}, & r_{\rm diff} < l_0, \\
1.9 \times 10^{-4} \left( \frac{\lambda_0}{1\,{\rm m}} \right)^{22/5} {\rm s}, & r_{\rm diff} > l_0 \\
\end{array} \right.
\end{eqnarray}
The estimated amplitude of this effect therefore depends on the effective scattering measure.  This equation furnishes a means to deduce the scattering measure of the IGM from measurements of temporally smeared impulsive events.

However, we caution that the effective scattering measure differs from the usual (Euclidean space) definition in terms of an integral of $C_N^2$ over path length in two respects. (i) The wavelength of radiation from a redshifted source is larger when it reaches the observer than when it was scattered by an object at high redshift, say at $z_L$, so that the observed SM is a factor $(1+z_L)^2$ smaller than the correct SM.  (ii) For scattering that occurs in the Hubble flow in which the Friedmann-Robertson-Walker metric applies (i.e. not in gravationally collapsed objects), the integral over path length $dl$ must be corrected for the curvature of spacetime and is replaced by an integral over redshift: $d_H (z) dz/(1+z)$.
\item For the purpose of completeness, we have attempted to estimate the expected effective scattering measure associated with the objects at cosmological distances.   A simple estimate suggests that the effective scattering measure is of the order of $5 \times 10^{13}\,$m$^{-17/3}$ at $z<1$ and $5 \times 10^{15}\,$m$^{-17/3}$ at ($z<10$) for the diffuse IGM. The contribution from collapsed systems such as intervening galaxies and cluster gas will dominate the effective scattering measure for sightlines through which they intersect, but is unlikely to occur; there is only a $5\%$ chance that a source at $z \lesssim 1.5$ and $z \lesssim 0.5$ will intersect a galaxy or a galaxy cluster, respectively. An extrapolation of the DM-SM relation observed in our Galaxy and modified to the densities applicable in the IGM yields an estimate of the expected SM that is also comparable to the two foregoing estimates.  

These estimates are compatible with existing limits on the angular broadening of compact sources at cosmological distances.  The deduced angular sizes of intra-day variable quasars only limit the effective scattering measure to ${\rm SM}_{\rm eff} < 10^{17}\,$m$^{-17/3}$ at $z \sim 1$.  

The absence of detected temporal smearing in several of the extragalactic transients detected by Keane et al. (2011) and Thornton et al. (2013) limits the effective scattering measure of the IGM to values below $\sim 5 \times 10^{15}\,$m$^{-17/3}$ at $z \la 1$ while the corresponding limit for the transient in which temporal smearing is detected is $\sim 2 \times 10^{16}\,$m$^{-17/3}$; this value is a limit in the sense that it is unclear whether the host galaxy makes a substantial additional contribution to the temporal smearing of the signal.

Lorimer et al.\,(2013) make use of an empirical DM-SM relation in their consideration of event rates for FRBs at frequencies relevant to the MWA and LOFAR.  They note that the amount of scattering observed in FRBs detected so far falls considerably below the empirical DM-SM relation observed in the Milky Way.  The rederivation of the DM-SM relation for the physical conditions of the IGM, in \S \ref{DMSM}, shows why this is the case.  One expects the amount of scattering expected per unit of dispersion measure to be several orders of magnitude lower than that found in the ISM.  Moreover, one predicts that the constant of proportionality between DM and SM to change with redshift as $(1+z)^3$.

\item These estimates of the scattering measure at $z \la 3$ suggest that temporal smearing may, {\it on average}, be less than $\sim 1\,$ms for observations at frequencies above 300\,MHz.   However, if the IGM is clumpy one may expect variations between differing lines of sight through the IGM in much the same way that large sight-line variations are observed in our Galaxy's turbulence ISM. 
\item The effects of scattering in the IGM and that in the host galaxy of a bright transient event are strongly distinguished by their dependence on redshift.  This is because the baryon density in the IGM increases sharply ($\propto (1+z)^3$) whereas the baryon density in the gravitationally collapsed systems (e.g. galaxies and clusters) is decoupled from the Hubble flow.  

For decoupled systems one has $\tau \propto (1+z_L)^{-3}$ if the diffractive scale is smaller than the inner scale of the turbulence and $\tau \propto (1+z)^{(2+\beta)/(2-\beta)}$ if it exceeds the inner scale;  one expects $\tau \propto (1+z)^{-17/5}$ if the turbulence follows the Kolmogorov value of $\beta=11/3$. 
In the IGM, the redshift dependence of the scattering depends on both the redshift dependence of the effective scattering measure and on the redshift depedence of the lever arm effect associated with the effective distance $D_{\rm eff}$.   We find $\tau_{\rm IGM} \sim z^2$ for $z \la 1$ and $(1+z)^{0.2-0.5}$ for $z \ga 1$.
\end{itemize}

\newpage

\appendix

\section{Summary of parameters used}
In Table \ref{SymTable}1 we present a summary of the most common symbols and parameters used throughout the text. \placetable{SymTable}

\begin{deluxetable}{l l l} 
\tabletypesize{\footnotesize}
\tablewidth{0pt}
\tablecaption{Table of Symbols \label{SymbolTable}} 
\tablehead{
\colhead{Symbol} & \colhead{Meaning} & First relevant equation
}
\startdata
$\beta$ & power-law index of the spectrum of turbulent densitity inhomogeneities & \ref{PhiDefn} \\
$C_N^2$ & Amplitude of the density power spectrum & \ref{PhiDefn} \\
$C_{N,{\rm gal}}^2$ & contribution to the density power spectrum from a galaxy & \ref{CN2gal} \\
$C_{N,{\rm icm}}^2$ & contribution to the density power spectrum from the intra-cluster medium & \ref{CN2icm} \\
$C_{N,{\rm obj}}^2$ & contribution from either $C_{N,{\rm gal}}^2$  or $C_{N,{\rm icm}}^2$  & \ref{SMicmgal} \\
$C_{N 0}^2$ & contribution to the density power spectrum from a turbulent clump in the IGM & \ref{SMclumpContribution} \\
$d_H$ & Hubble radius & \ref{dHubble} \\
$D_{LS}$ & Angular diameter distance between the scattering region and the source & \ref{rF} \\
$D_L$  & Angular diameter distance between the observer and the scattering region & \ref{rF} \\
$D_S$ & Angular diameter distance between the observer and the source & \ref{rF} \\
$D_{\rm eff}$ & $D_L D_{LS}/D_S$ & \ref{tauEmpirical} \\
$\Delta L$ & Scattering region depth & \ref{DphiDefn} \\
$F_c$ & turbulence fluctuation parameter & \ref{FcDefn} \\
$k$ & wavenumber &  \ref{thetaApprox} \\
$\lambda_0$ & wavelength in the frame of the observer & \ref{rF} \\
$l_0$ & Turbulence inner scale & \ref{PhiDefn} \\
$L_0$ & Turbulence outer scale & \ref{PhiDefn} \\
$r_{\rm diff}$ & Diffractive scale length & \ref{rdiffDefn} \\
$r_{\rm F}$  & Fresnel scale, $D_L D_{LS} \lambda_0/(2 \pi D_S (1+z_L)$ & \ref{rF}\\
SM & Scattering measure & \ref{rdiffDefn} \\
SM$_{\rm eff}$  & Effective (redshift-corrected) scattering measure & \ref{SM0eq} \\
$\tau$ & Temporal broadening time & \ref{tauApprox} \\
$\tau_{\rm IGM}$ & Temporal broadening time associated with IGM scattering & \ref{tauIGMLowHighZ}\\
$\tau_{\rm host}$ & Temporal broadening time associated with scattering in the host galaxy & 
\ref{tauhost} \\
$z_L$ & redshift of the scattering region & \ref{rF}  \\
$z_S$ & redshift of the transient event (i.e.\,source) & \ref{tauIGMLowHighZ} \\
\enddata
\end{deluxetable} \label{SymTable}

\newpage 
\section{The contribution from a clumpy IGM} \label{ClumpApp}
In this appendix we consider in detail the contribution of a collection of clouds embedded in the IGM whose density fluctuations follow a log-normal probability distribution.  The form of the distribution is, 
\begin{eqnarray}
p(x=n_e) = \frac{1}{\sqrt{2 \pi} x \sigma} \exp \left[ - \frac{(\mu - \ln x)^2}{2 \sigma^2} \right], \quad x >0,
\end{eqnarray}
where $\mu$ is the mean density and $\sigma$ is the standard deviation of the associated normal distribution.  The mean of a lognormal distribution is $\exp(\mu + \sigma^2/2)$ and its variance is $\exp(2 \mu +\sigma^2) (e^{\sigma^2} -1)$, so if the root-mean-square density is a factor $f$ times the mean density, and the mean is $\mu_0$, one has
\begin{eqnarray}
\mu &=& \ln \mu_0 - \frac{1}{2} \ln(f^2+1), \\
\sigma &=& \sqrt{\ln(f^2+1)}. 
\end{eqnarray}
Thus the corresponding probability of obtaining a density greater than $N_e$ is,
\begin{eqnarray}
P(X > N_e) 
&=& 1 - \frac{1}{2} {\rm erfc} \left[ \frac{\ln \mu_0 - \frac{1}{2} \ln(f^2+1) -\ln X}{\sqrt{2 \ln (f^2+1)}} \right]. 
\end{eqnarray}
This probability indicates the chance of obtaining an overdensity greater than some value $N_e$ in a given cell of turbulence.  Now, if the turbulence has an outer scale $L_0$ the fluctuations between adjacent cells of length $L_0$ can be taken to be mutually independent, and thus each $L_0^3$-sized volume in the IGM contains an independent realization of the stochastic turbulence.  For any wavefront propagating through a path length $L$ through the IGM, the radiation will encounter $\sim L/L_0$ independent turbulent cells.  

We can use these arguments to determine the rate per unit length of encountering a region of overdensity $N_e$: if the probability per unit cell is $P$, then the rate per unit length of encountering an overdensity region is $P/L_0$.  Now, in general $P$ will depend on the redshift, because the mean density of the IGM scales as $(1+z)^3$.  We therefore write the mean density as $\mu_0 (1+z)^3$ and, since the variance is a multiple $f^2$ of the mean-squared, it follows that the standard deviation increases like $ \mu_0 (1+z)^3 f$.  Other scalings of $\sigma$ with redshift may be feasible, but we expect $\sigma$ to depend on the microphysics of the turbulence, which should be redshift-independent, so we adopt the foregoing assumption as a conservative approach.  So we can now compute the mean number of overdensitites that the radiation will ecounter as,
\begin{eqnarray}
N= \int_0^L \frac{P(z)}{L_0} dl = \int_0^{L/c} \frac{P(z)}{L_0} c dt.
\end{eqnarray}
One thus obtains
\begin{eqnarray}
N (X>N_e;z)
&=& \frac{c }{L_0 H_0} \int_0^z \frac{  1 - \frac{1}{2} {\rm erfc} \left[ \frac{\ln [\mu_0 (1+z)^3] - \frac{1}{2} \ln(f^2+1) -\ln X}{\sqrt{2 \ln (f^2+1)}} \right] }{(1+z)\sqrt{ \Omega_\Lambda + \Omega_m (1+z)^3}} dz. \label{NumberBlobsApp}
\end{eqnarray}
A value of $N$ much smaller than one signifies that it is improbable that a line of sight up to a redshift $z$ intersects any clouds above a density $X=N_e$.  

\acknowledgments
Parts of this research were conducted by the Australian Research Council Centre of Excellence for All-sky Astrophysics (CAASTRO), through project number CE110001020.


\bibliographystyle{apj}

\end{document}